\documentclass[twocolumn,showpacs,amsmath,amssymb]{revtex4}

\usepackage{graphicx}   % need for figures
\usepackage{subfigure}  % use for side-by-side figures
\usepackage{dcolumn}    % Align table columns on decimal point
\usepackage{bm}         % bold math

\begin{document}

\title{Fitting Single Particle Energies in $sdgh$ Major Shell}

\author{Erdal Dikmen}
\affiliation{ Department of Physics, S\"{u}leyman Demirel University,
32260 Isparta, Turkey}
\author{O\u{g}uz \"{O}zt\"{u}rk}
\affiliation{ Department of Physics, S\"{u}leyman Demirel University,
32260 Isparta, Turkey}
\author{Yavuz Cengiz}
\affiliation{ Department of Electronics and Communication Engineering, 
S\"{u}leyman Demirel University, 32260 Isparta, Turkey}

\date{\today}

\begin{abstract}
We have performed two kinds of non-linear fitting procedures to the single-particle 
energies in the $sdgh$ major shell to obtain better shell model results.
The low-lying energy eigenvalues of the light Sn isotopes with $A=103-110$ 
in the $sdgh$-shell are calculated in the framework of the nuclear shell model 
by using CD-Bonn two-body effective nucleon-nucleon interaction. The obtained 
energy eigenvalues are fitted to the corresponding experimental values by using 
two different non-linear fitting procedures, i.e., downhill simplex method and 
clonal selection method. The unknown single-particle energies of the states 
$2s_{1/2}$, $1d_{3/2}$, and $0h_{11/2}$ are used in the fitting methods to obtain 
better spectra of the $^{104,106,108,110}$Sn isotopes. We compare the energy 
spectra of the $^{104,106,108,110}$Sn and $^{103,105,107,109}$Sn isotopes with/without 
a nonlinear fit to the experimental results.

\end{abstract}

% PACS, the Physics and Astronomy
                                   % Classification Scheme.
\maketitle

\section{Introduction}

The nuclear shell model assumes any nuclei as a many-body quantum mechanical system
specified by the Hamiltonian in the second quantization
\begin{equation}
H=\sum \epsilon _{\alpha }a^{\dag}_{\alpha }a_{\alpha }+\sum 
                 V_{\alpha \beta \gamma \delta }^{eff}
                 a^{\dag}_{\alpha }a^{\dag}_{\beta }a_{\delta }a_{\gamma }
\label{equ:many-body_H}
\end{equation}
The solution of this Hamiltonian for a series of nuclei in the periodic table 
implies deriving systematically some nuclear properties such as the energy 
spectra, electromagnetic transitions, and beta decay, etc \cite{Talmi-93,Heyde-94}.

The quantum many-body solution of the Hamiltonian in Eq. (\ref{equ:many-body_H})
for a series of nuclei is very complex that is normally divided in two 
sub-problems: the first one being the establishment of an effective interaction
$V^{eff}$ from the bare nucleon-nucleon (NN) potential in a valence space and 
of a set of valence space single-particle energies for sequences of nuclei. 
The second problem is the quantum many-body calculations of the Hamiltonian 
in Eq. (\ref{equ:many-body_H}) \cite{Dikmen-06_1}. In this article, we present 
an improvement under the first problem.

The one-body part of the Hamiltonian in Eq. (\ref{equ:many-body_H}), i.e., the first 
term, describes the interaction of the valence nucleons with the closed core. The 
second term is the two-body part which describes the two-body effective interaction. 
In this study we have used the CD-Bonn two-body effective interactions in
the $sdgh$-shell provided to us by M. Hjorth-Jensen et al. \cite{Jensen-98}. 
They calculate the charge-dependent Bonn (CD-Bonn) two-body effective interaction 
$V^{eff}$ for shell model studies based on the free nucleon-nucleon interaction 
$V$ \cite{Jensen-95,Machleidt-96,Machleidt-01}. However, the single-particle energies 
can not be obtained via their procedure. The latter are normally taken from the 
experimental data on the nuclei one particle away from the closed shell nuclei. 
Unfortunately, in the $sdgh$ major shell, the nuclei one particle away from the 
closed shell nucleus $^{100}$Sn have so far escaped by measurements. Therefore, 
we used instead the single-particle energies for neutrons in the $sdgh$ major 
shell in our earlier studies around $A=100$ mass region
\cite{Dikmen-01,Dikmen-02,Dikmen-06_2,Dikmen-07,Dikmen-09_1,Dikmen-09_2}, e.g.,
$\varepsilon _{1d_{5/2}}=0.00$ MeV, $\varepsilon _{0g_{7/2}}=0.08$ MeV,
$\varepsilon _{2s_{1/2}}=2.45$ MeV, $\varepsilon _{1d_{3/2}}=2.55$ MeV, and
$\varepsilon _{0h_{11/2}}=3.20$ MeV. These energies are in reasonable agreement 
with similar shell model calculations in this region, see for example Ref. 
\cite{Andreozzo-97,Engeland-00, Engeland-02}. However, in the recent experimental 
work of Ref. \cite{Seweryniak-07} an excited state at 171.7 keV has been 
identified in $^{101}$Sn and interpreted as the $0g_{7/2}$ single-neutron state.

The experiment in Ref. \cite{Seweryniak-07} fixes the ambiguity about the 
single-particle energy of the $0g_{7/2}$ orbit. So, the uncertain single-particle 
energies in the $sdgh$ major shell are left only on three of them, namely 
$\varepsilon _{2s_{1/2}}$, $\varepsilon _{1d_{3/2}}$, and $\varepsilon _{0h_{11/2}}$. 
The ambiguity of these single-particle energies establishes the need of this work.
In this work, we have performed two kinds of non-linear fitting methods, namely
downhill simplex method and clonal selection principle, to obtain
the optimum single-particle energies for the single-particle orbits of $2s_{1/2}$, 
$1d_{3/2}$, and $0h_{11/2}$.

In order to obtain a better energy spectrum of any isotopes in $sdgh$ major shell 
in the nuclear shell model framework we should improve both the two-body effective 
interactions and the single-particle energies given in Eq. (\ref{equ:many-body_H}). 
Wildenthal in Ref. \cite{Wildenthal-84} provided a fit procedure in the $sd$-shell 
whereby be improved on an effective interaction by fitting the 3 single-particle 
energies and 63 effective two-body matrix elements. The question being considered 
in that study is: What is the best possible shell model description of the nuclei 
in the $sd$-shell independently of what the Hamiltonian is? The same ambitious 
question was raised by Honma \textit{et al.} \cite{Honma-02} for the $pf$-shell, 
where there are 4 single-particle energies and 195 effective two-body matrix 
elements to fit.

In the $sdgh$-shell there are 5 single-particle energies and 542 two-body matrix 
elements of effective interaction $V^{eff}$ that specifies the Hamiltonian. 
A global fit of both the single-particle energies and these two-body matrix 
elements is certainly out of question. Many of the $sdgh$-shell nuclei can not 
be calculated yet, some of the nuclei that can be modeled require very large 
calculations and we have limited experimental data. This study restricts itself 
to fit single-particle energies.

We have five single-particle states in the $sdgh$ major shell. All theoretical 
and experimental data indicate that the lowest single-particle state is the 
$1d_{5/2}$ state and the energy of the $0g_{7/2}$ state are experimentally known 
\cite{Jensen-98,Jensen-95,Dikmen-01,Dikmen-02,Dikmen-06_2,Dikmen-07,Dikmen-09_1,
Dikmen-09_2,Andreozzo-97,Engeland-00,Engeland-02,Seweryniak-07}. So we fix 
the $1d_{5/2}$ state as the lowest at the energy of 0.00 MeV and the $0g_{7/2}$ 
state as the first excited state at the energy of 0.172 MeV, and then perform a 
non-linear fit to the other three single-particle energies. We use the downhill 
simplex method \cite{simplex} and the clonal selection principle 
\cite{clonal-selection} in three dimensions to fit these three unknown 
single-particle energies. The downhill simplex method was chosen over the others, 
i.e., steepest decent annealing, monte carlo, because of the claim that it avoids 
trapping in local minima. The clonal selection principle was chosen because
it represents a global fit over the all cells.

\section{Fitting Single Particle Energies in 3-D}

Because of the complexity and very large model space of the two-body effective
interaction in the $sdgh$ major shell, we only apply a fit process to the values 
of the single-particle energies. We adjust the values of the single-particle 
energies so as to fit the theoretical energy levels of the selected nuclei to 
the experimental energy levels. We outline some major steps in the present study
and the fitting procedures below.

For a set of known experimental energy data $E^J_{exp}(J= J_1,J_2,...)$ with the 
angular momentum $J$ of a specific nucleus with a mass $A$, we calculate corresponding 
shell-model energy eigenvalues $E_J^{theory}$ . We minimize the quantity
\begin{equation}
\sigma^2 = \sum_A \sum_{J=J_1,J_2,...} (E^J_{exp} - E^J_{theory})^2 
\label{equ:sigma}
\end{equation}
by varying the values of the single-particle energies. Since the fitting methods 
that we use, simplex and clonal selection, are the non-linear processes with respect 
to the parameters, it is solved in an iterative way with successive variations of 
those parameters (single-particle energies) followed by diagonalization of the 
Hamiltonian until convergence.

We choose the light even-even Tin isotopes experimentally available. Since the fit 
process is a non-linear process, the selection of the heavier nuclei would require 
very long calculation times until convergence. Therefore the nuclei 
$^{104,106,108,110}$Sn are the largest Tin isotopes for which we can fit the 
single-particle energies in some reasonable time. On the other hand, the more number
of nuclei are included in the fit process, the more accurate fitted single-particle 
energies are obtained in the $sdgh$ major shell.

There are 5 single-particle energies in $sdgh$-shell model space. For simplicity 
we reduce the dimension of the problem to 3-D instead of 5-D by fixing the energies 
of the $1d_{5/2}$ state to be 0.000 MeV and the $0g_{7/2}$ state to be 0.172 MeV. 
The energy values of the other three single-particle states $2s_{1/2}$, $1d_{3/2}$, 
and $0h_{11/2}$ are varied during fitting process. The experimental energies used for 
the fit are limited to those of the first occurrences of low-lying states. The 
theoretical energies are also chosen to be the lowest in the theoretical spectra.

\subsection{The Downhill Simplex Method in Multidimensions}

The downhill simplex method is a geometrical hill climbing scheme to find the
minimum of a function of more than one independent variable. The method requires
only function evaluations. A simplex is a geometrical figure with $N+1$ points
(or vertices) that lives in a parameter space of dimension $N$ and all
interconnecting line segments and polygonal faces of vertices. In 2-D space a
simplex is a triangle, in 3-D space it is a tetrahedron, and so on. Given the
function value at each of the simplex's vertices, the worst vertex is displaced
by having the simplex undergo one of four possible ``moves", namely
reflection, reflection and expansion, contraction, or multiple contraction
(see more details in Ref. \cite{simplex,num-recipes}).

The downhill simplex method starts with $N+1$ points defining an initial simplex.
If one of $N+1$ points is the initial starting point $P_0$, then the other $N$
points are
\begin{equation}
P_i=P_0 + \lambda e_i 
\label{equ:simplex}
\end{equation}
where $e_i$'s are $N$ unit vectors, and $\lambda$ is a constant characterizing the
problem's length scale ( or $\lambda$ may be different on each direction).

The downhill simplex method takes a series of moves, most moves just moving
the point of the simplex where the function is largest (``highest point") through
the opposite face of the simplex to a lower point. These moves are called reflections,
and they are constructed to conserve the volume of the simplex. When it can do so,
the method expands the simplex in one or another direction to take larger moves. 
When it reaches a ``valley floor" the simplex contracts itself in the transverse 
direction and tries oozing down the valley. If there is a situation where the simplex 
is trying to ``pass through the eye of a needle" it contracts itself in all directions, 
pulling itself in around its lowest (best) point. The simplex undergoes successive
such moves until no move can be found that leads to further improvement beyond some
present tolerance.

The simplex method requires that one provides initial coordinates (x,y) for 
the simplex's four vertices. Despite the simplex method's pseudo-global abilities 
on a multidimensional and global problem, the choice of the initial location for the
simplex often determines whether the global maximum or minimum is ultimately found.

\subsection{The Clonal Selection Principle}

Artificial Immune Systems (AIS) are computational methods inspired by biological 
principles of the natural immune system. AIS uses the ideas from immunology to 
develop an algorithm describing adaptive systems in various areas of science 
and engineering applications \cite{Bernardino-09,deCastro-02}. The clonal selection
algorithm is based on the idea of AIS and focuses on a systematic view of the immune
system. The algorithm is initially proposed to carry out machine-learning and 
pattern-recognition tasks and then is applied to optimization problems 
\cite{deCastro-02}. In the optimization problems, this algorithm uses random 
instead of certain passing rules to avoid sticking on a local minima.

Clonal selection algorithm uses two kinds of populations of strings: a set of 
antigens (corresponding to a objective function) and a set of antibodies 
(corresponding to variables of the objective function) within the maximum and 
minimum limits of variables. Antibody population is generated by random numbers 
scaled by limits of variables. The affinity value of an antibody matches the value 
of purpose function calculated for a given antibody. The affinity population matrix 
is obtained by sorting affinity values. The amount of antibodies having highest
affinity are selected from the entire population and a new set of population is
formed as a sub-antibody array. 

A cloning ratio is determined by a percentage of each antibody array elements. 
Each of antibody array elements is cloned by 
cloning ratio and inverting cloning factor determining the number of cloning.
The affinity values of the new cloned population exposed to mutation are 
calculated and then a hypermutation process is performed by inverse cloning 
ratio. After the hypermutation process, a new number of antibody are determined
and the cloned antibody array is obtained by sorting the corresponding affinity 
values. An amount of antibodies with highest affinity are reselected and added 
to the antibody population set. Finally, an amount of antibodies with low affinity 
value within the antibody population set are replaced with the newly formed 
antibodies. At the end, the optimum values are determined.

\section{Results and Discussion}

The initial values of the single-particle energies to be fitted are chosen to be 
$\varepsilon _{2s_{1/2}}=2.45$ MeV, $\varepsilon _{1d_{3/2}}=2.55$ MeV, and
$\varepsilon _{0h_{11/2}}=3.20$ MeV. These energies are in reasonable agreement
with similar shell model calculations in this region \cite{Dikmen-01,Dikmen-02,
Dikmen-06_2,Dikmen-07,Dikmen-09_1,Dikmen-09_2,Andreozzo-97,Engeland-00,Engeland-02}.

The calculation time for each iteration in both simplex and clonal selection 
fitting methods is very long time since it requires the shell-model calculations 
of the states $J^{\pi}=0^+,2^+,4^+,6^+,8^+,10^+$ for the Tin isotopes 
$^{104,106,108,110}$Sn. We have carried out the shell-model calculations in 
the High Performance Parallel Computer Cluster at Suleyman Demirel University, 
Turkey, consisting of 1 master and 30 slave nodes with each one of 3.2 GHz Dual 
Xeon Processors. For a comparison, the required time to calculate the energy of 
the state $4^+$ of $^{108}$Sn and $^{110}$Sn are about 10 min. and 156 min., 
respectively. The largest Hamiltonian dimension among the states that we include 
in the fitting process is 156674 for the state $8^+$ of $^{110}$Sn.

%%%%%%%%%%%%%%%%%%%%%%%%%%%%%%%FIG1%%%%%%%%%%%%%%%%%%%%
\begin{figure}[t]
\includegraphics[scale = 0.48]{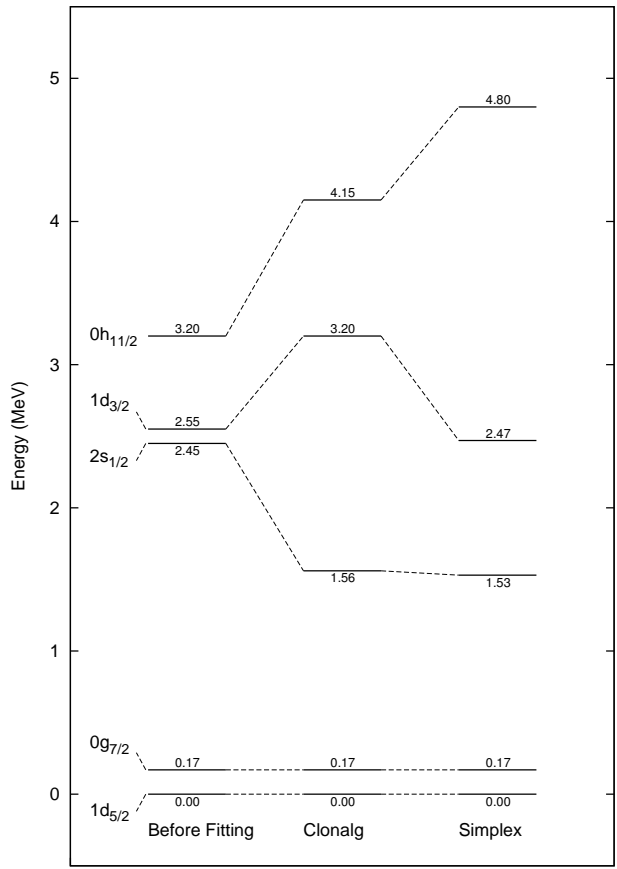}%
\caption{The single-particle energies before and after the fit processes.}
\label{fig:spe}
\end{figure}
%%%%%%%%%%%%%%%%%%%%%%%%%%%%%%%%%%%%%%%%%%%%%%%%%%%%%%% 

%%%%%%%%%%%%%%%%%%%%%%%%%%%%%%%FIG2%%%%%%%%%%%%%%%%%%%%
\begin{figure}[!ht]
%\centering
\subfigure{\includegraphics[scale=0.41]{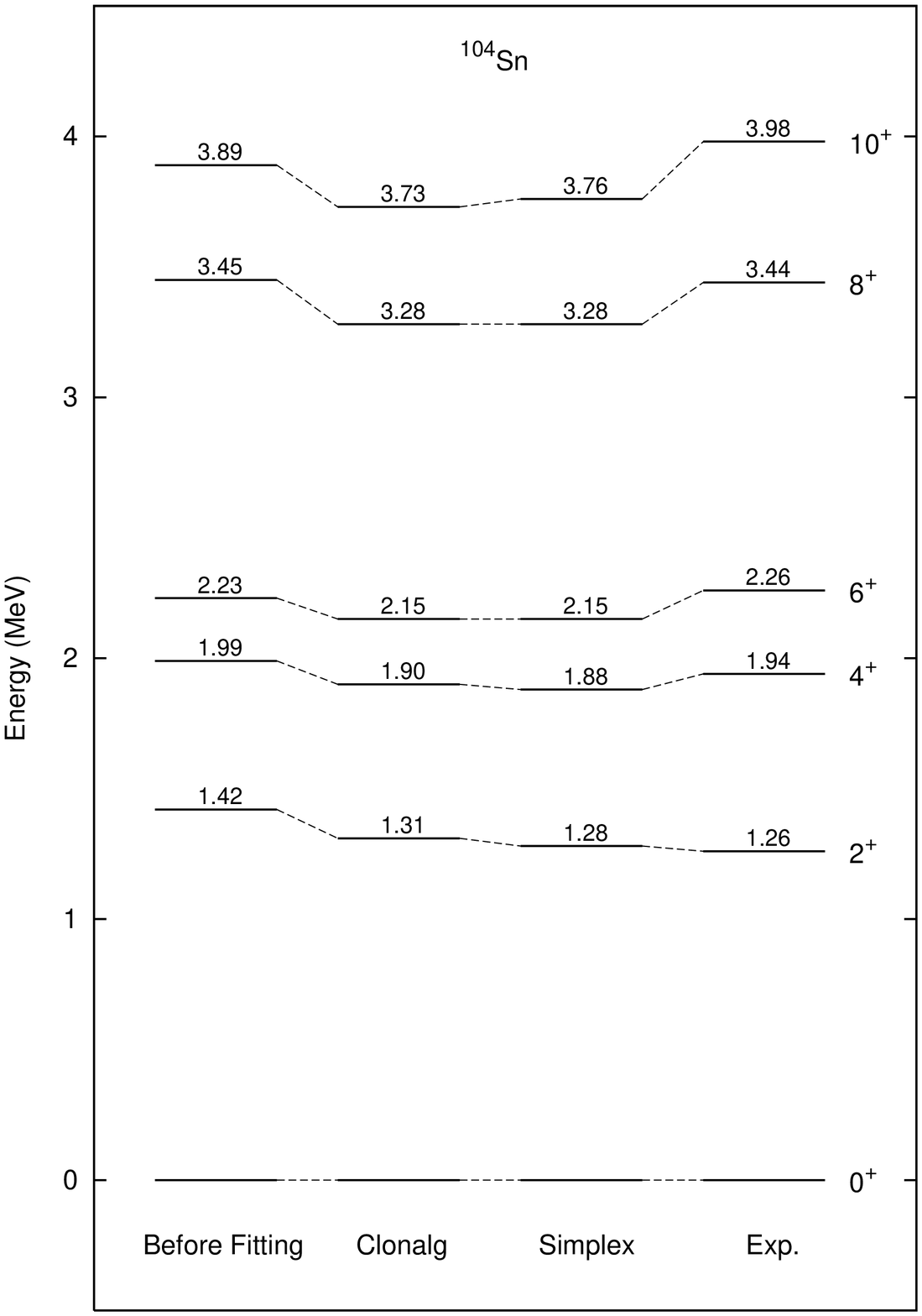}}
\subfigure{\includegraphics[scale=0.41]{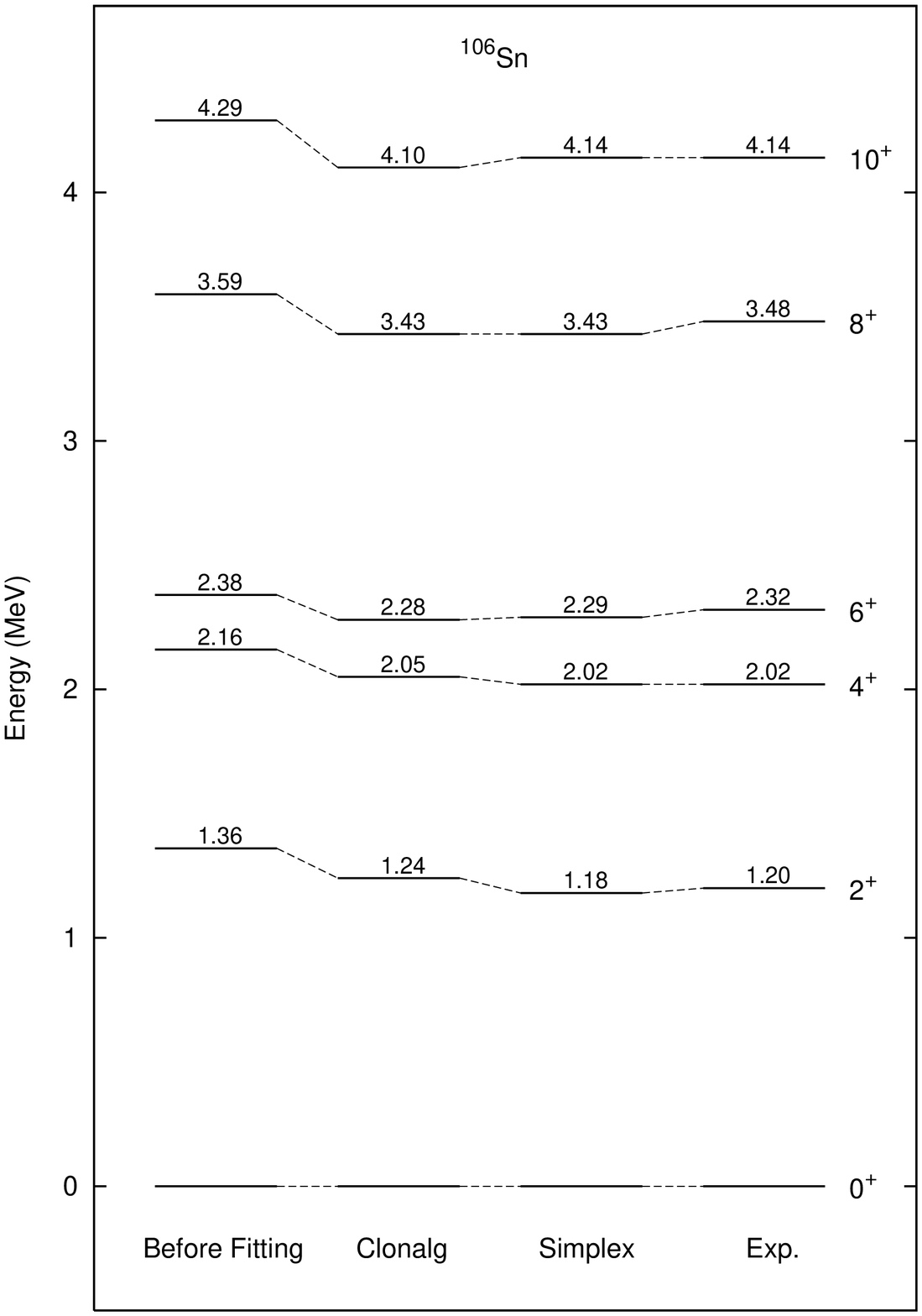}}
\caption{Experimental and theoretical spectrum of $^{104}$Sn and $^{106}$Sn
obtained with the single-particle energies before fit and the fitted ones
using the CD-Bonn effective interaction.}
\label{fig:Sn104_106}
\end{figure}
%%%%%%%%%%%%%%%%%%%%%%%%%%%%%%%FIG3%%%%%%%%%%%%%%%%%%%%
\begin{figure}[!ht]
\centering
\subfigure{\includegraphics[scale=0.41]{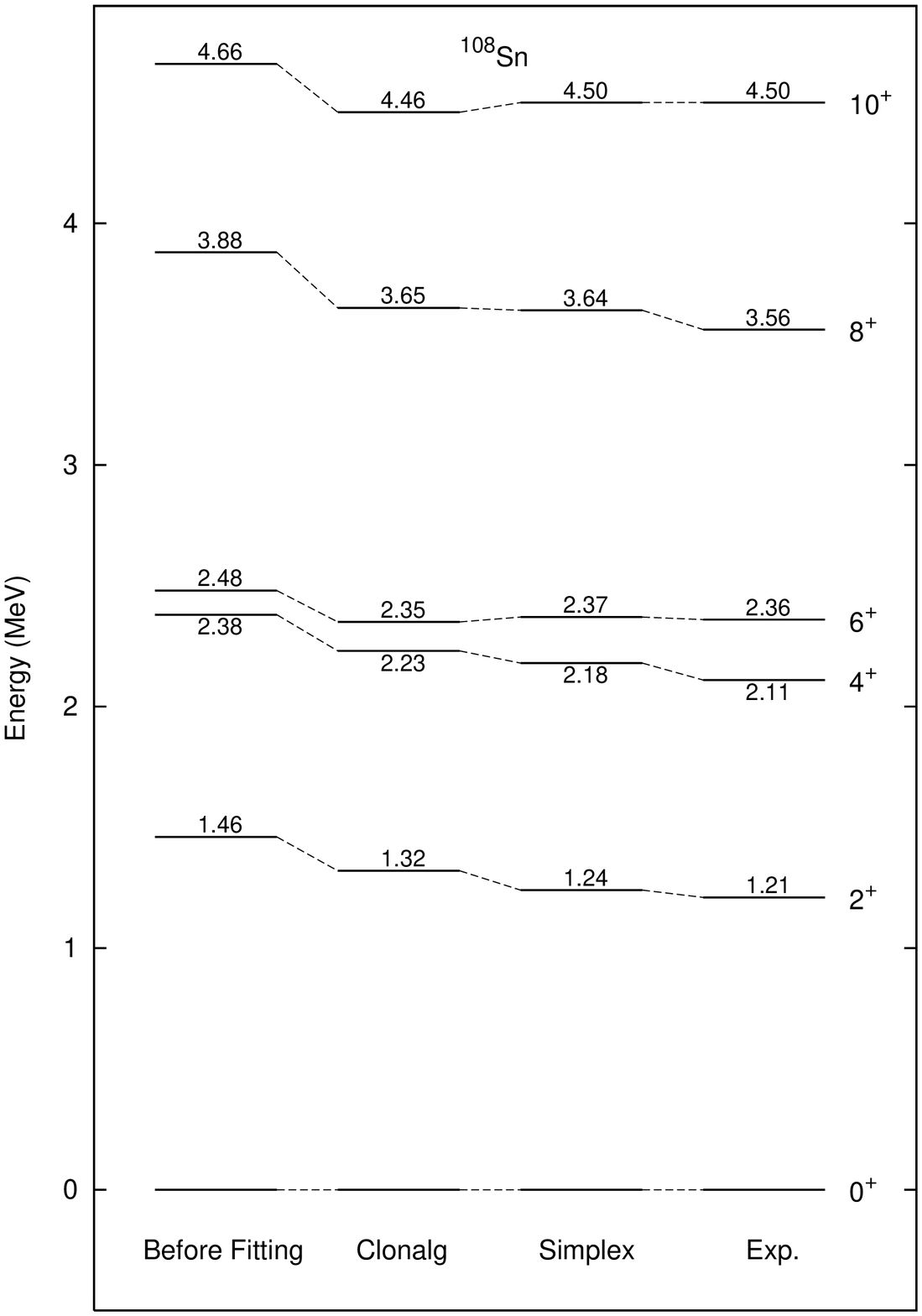}}
\subfigure{\includegraphics[scale=0.41]{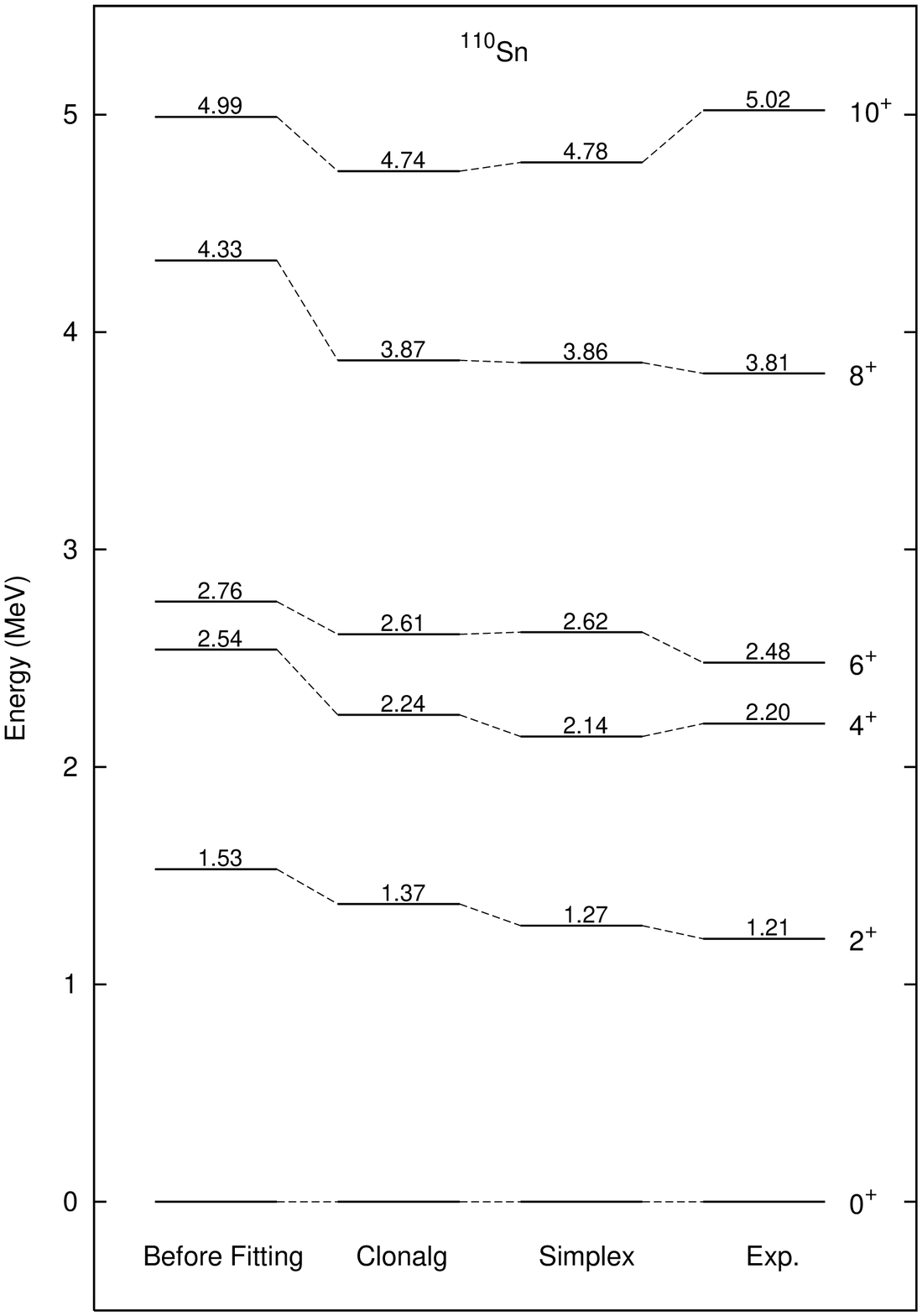}}
\caption{Experimental and theoretical spectrum of $^{108}$Sn and $^{110}$Sn
obtained with the single-particle energies before fit and the fitted ones
using the CD-Bonn effective interaction.}
\label{fig:Sn108_110}
\end{figure}
%%%%%%%%%%%%%%%%%%%%%%%%%%%%%%%%%%%%%%%%%%%%%%%%%%%%%%% 

%%%%%%%%%%%%%%%%%%%%%%%%%%%%%%%FIG4%%%%%%%%%%%%%%%%%%%%
\begin{figure}[!ht]
\centering
\subfigure{\includegraphics[scale=0.41]{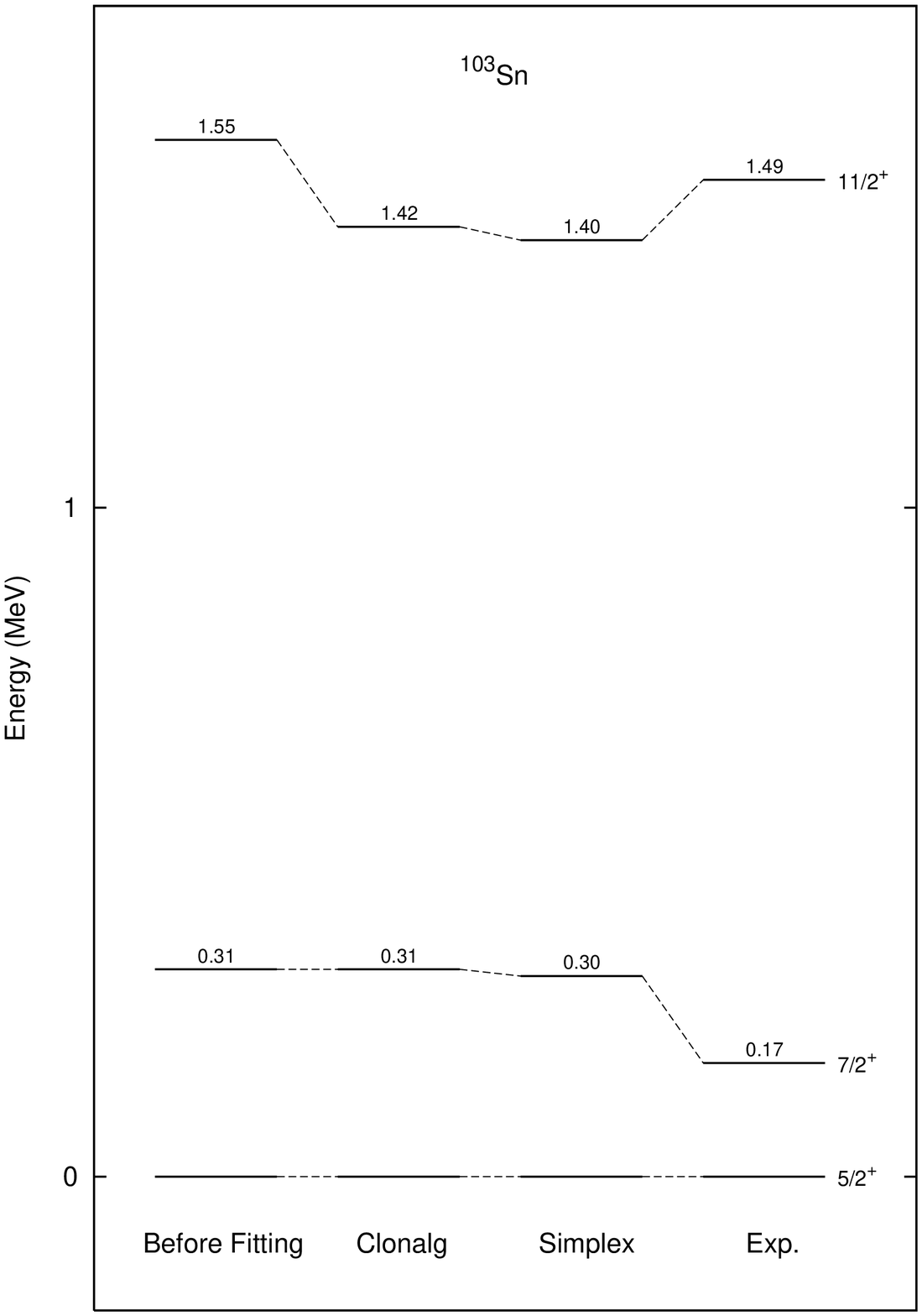}}
\subfigure{\includegraphics[scale=0.41]{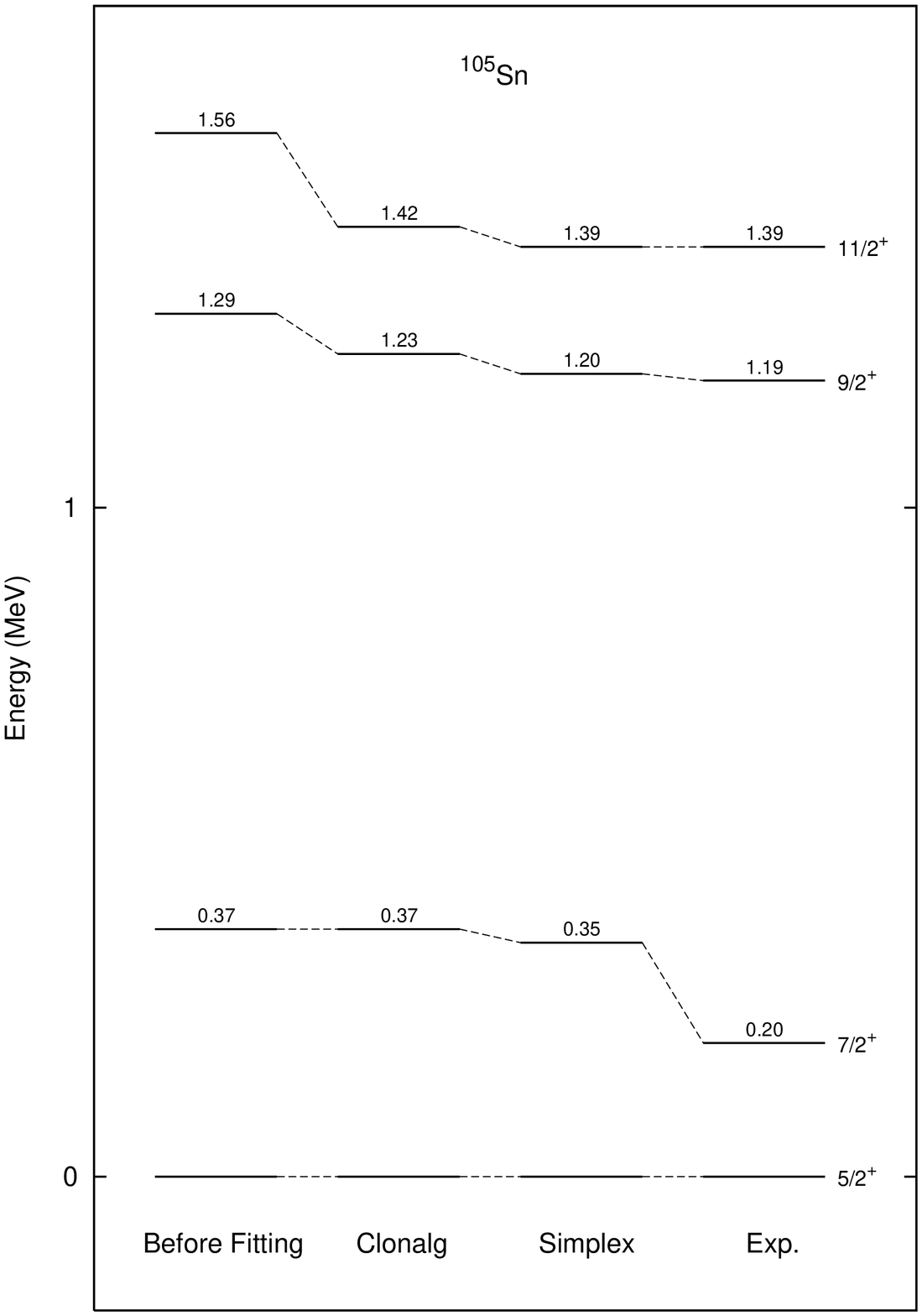}}
\caption{Experimental and theoretical spectrum of $^{103}$Sn and $^{105}$Sn
obtained with the single-particle energies before fit and the fitted ones
using the CD-Bonn effective interaction.}
\label{fig:Sn103_105}
\end{figure}
%%%%%%%%%%%%%%%%%%%%%%%%%%%%%%%FIG5%%%%%%%%%%%%%%%%%%%%
\begin{figure}[!ht]
\centering
\subfigure{\includegraphics[scale=0.41]{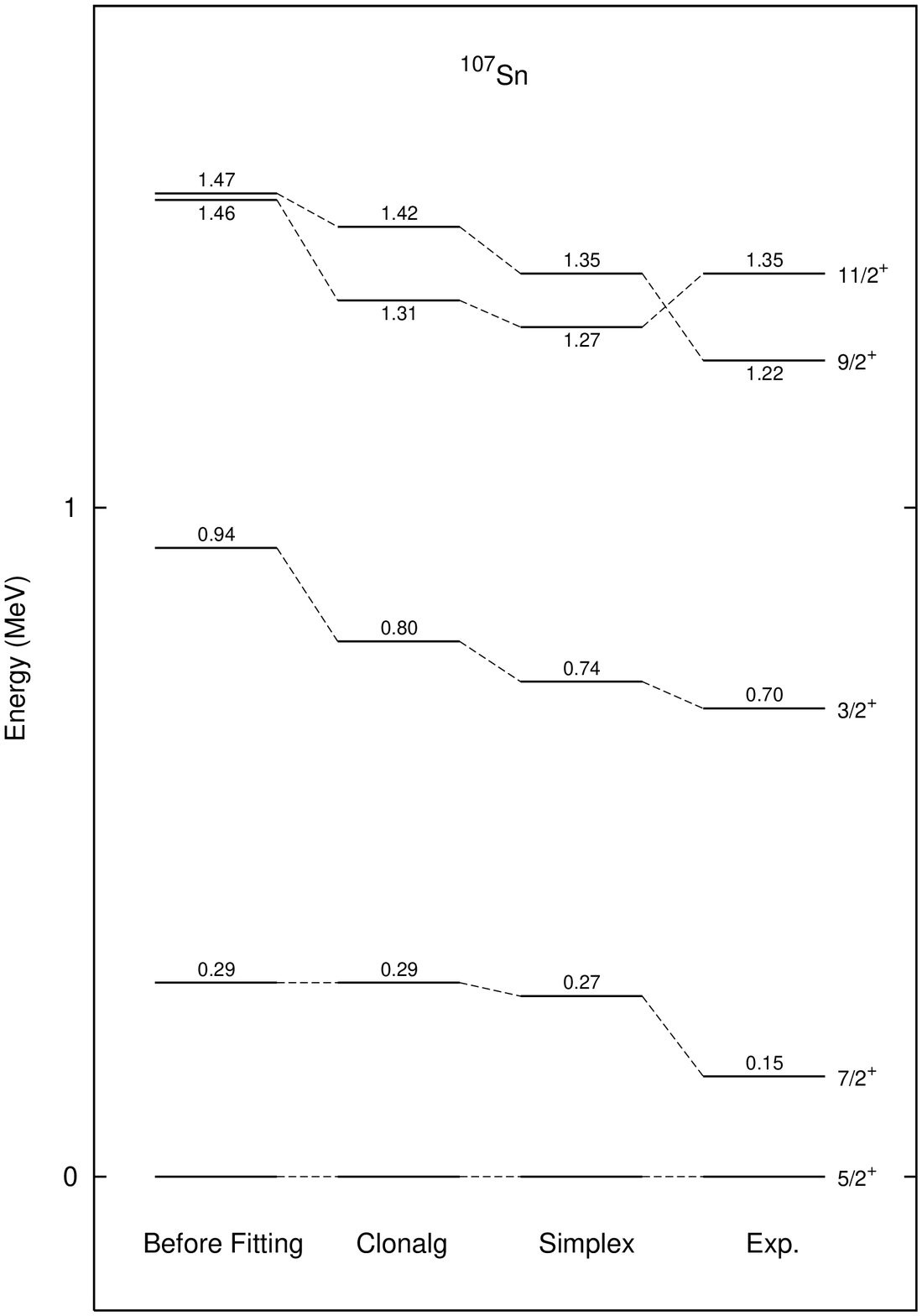}}
\subfigure{\includegraphics[scale=0.41]{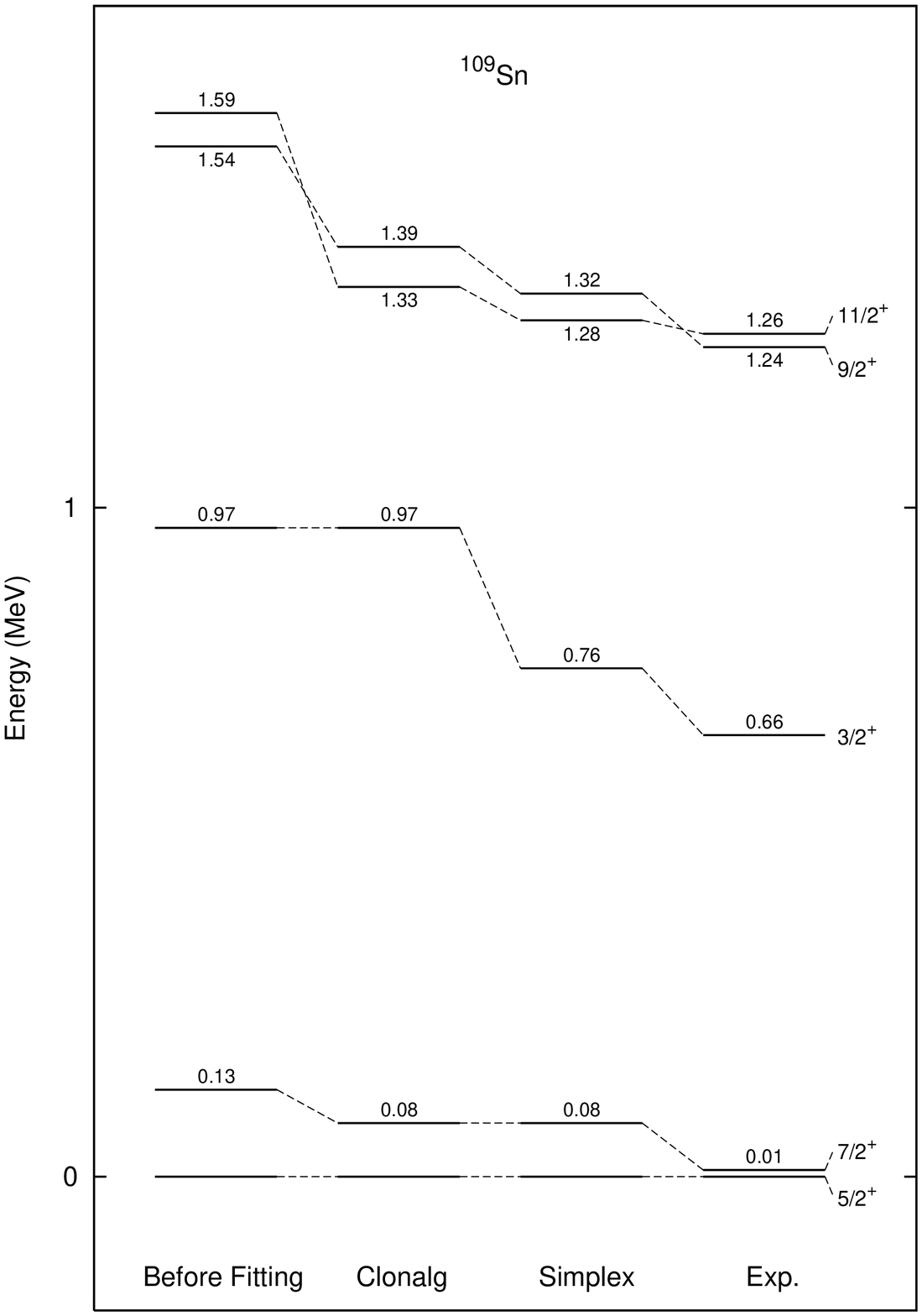}}
\caption{Experimental and theoretical spectrum of $^{107}$Sn and $^{109}$Sn
obtained with the single-particle energies before fit and the fitted ones
using the CD-Bonn effective interaction.}
\label{fig:Sn107_109}
\end{figure}
%%%%%%%%%%%%%%%%%%%%%%%%%%%%%%%%%%%%%%%%%%%%%%%%%%%%%%% 

\begin{table}
\caption{\label{tab:rms}The $\sigma$, root mean square (rms), values for the $2^+$, 
$4^+$, $6^+$, $8^+$, $10^+$ states of Sn isotopes with $A=104$-$110$ before and after 
fitting processes.}
\begin{ruledtabular}
\begin{tabular}{c|cccc}
$^{A}$Sn & $J^\pi_i$ & $\sigma_{\rm before}$ & $\sigma_{\rm clonal}$ & $\sigma_{\rm simplex}$  \\ \hline
           & $2^+$  & 0.16  & 0.05 & 0.02  \\
           & $4^+$  & 0.05  & 0.04 & 0.06  \\
$^{104}$Sn & $6^+$  & 0.03  & 0.11 & 0.11  \\
           & $8^+$  & 0.01  & 0.16 & 0.16  \\
           & $10^+$ & 0.09  & 0.25 & 0.22  \\ \hline
           & $2^+$  & 0.16  & 0.04 & 0.02  \\
           & $4^+$  & 0.14  & 0.03 & 0.00  \\
$^{106}$Sn & $6^+$  & 0.06  & 0.04 & 0.03  \\
           & $8^+$  & 0.11  & 0.05 & 0.05  \\
           & $10^+$ & 0.15  & 0.04 & 0.00  \\ \hline
           & $2^+$  & 0.25  & 0.11 & 0.03  \\
           & $4^+$  & 0.27  & 0.12 & 0.07  \\
$^{108}$Sn & $6^+$  & 0.12  & 0.01 & 0.01  \\
           & $8^+$  & 0.32  & 0.09 & 0.08  \\
           & $10^+$ & 0.16  & 0.04 & 0.00  \\ \hline
           & $2^+$  & 0.32  & 0.16 & 0.06  \\
           & $4^+$  & 0.34  & 0.04 & 0.06  \\
$^{110}$Sn & $6^+$  & 0.28  & 0.13 & 0.14  \\
           & $8^+$  & 0.52  & 0.06 & 0.05  \\
           & $10^+$ & 0.03  & 0.28 & 0.24  \\ 
\end{tabular}
\end{ruledtabular}
\end{table}
%%%%%%%%%%%%%%%%%%%%%%%%%%%%%%%%%%%%%%%%%%%%%%%%%%%%%%%%%%%%%

The fitted single-particle energies obtained by using clonal selection algorithm and 
simplex algorithm are shown in Figure \ref{fig:spe} as compared to the ones before 
fitting. As mentioned above, the energies of the single-particle states $1d_{5/2}$ 
and $0g_{7/2}$ are fixed to the values of 0.00 MeV and 0.17 MeV, respectively. 
Both fits give almost same fitted energy of the single-particle state $2s_{1/2}$
to be 1.56 MeV and 1.53 MeV by lowering almost 1 MeV from the initial value of
2.45 MeV. The energies of the other two single-particle states $1d_{3/2}$ and
$0h_{11/2}$ are very different each other after the fit processes. The simplex
method changes the energy of the single-particle state $1d_{3/2}$ -0.008 MeV as
to be 2.47 from the initial value of 2.55 MeV while clonal selection method makes
the change +0.65 MeV as to be 3.20 MeV. The major change happens to the energy of
the single-particle state $0h_{11/2}$: with a change of +0.95 MeV and +1.6 MeV
along with clonal selection and simplex methods, respectively.

As we mentioned in Section II, the convergence of optimization algorithms are 
satisfied by minimizing the quantity, $\sigma^2$, given in Eq. (\ref{equ:sigma}).
The value of $\sigma^2$ before fitting process is 1.127. The values of $\sigma^2$ 
after fitting processes are 0.208 after 76 iterations for simplex algorithm and 0.256 
after 130 iterations for clonal selection algorithm. Both simplex and clonal selection 
algorithms quit the iterations by having unchange value of $\sigma^2$, 0.208 and 0.256 
respectively. Therefore, the convergence of the fitted single-particle energies are 
obtained by lowering the value of $\sigma^2$ from 1.127 to 0.208 and 0.256 for simplex 
and clonal selection methods, respectively.

After having the fitted single-particle energies, we have carried out the shell-model
calculations for the states $0^+$, $2^+$, $4^+$, $6^+$, $8^+$, $10^+$ of even-even 
$^{104,106,108,110}$Sn isotopes to see how good spectrum obtained by using the new 
single-particle energies. In Table \ref{tab:rms}, we have given the root mean square 
(rms) deviations, $\sigma$, i.e., calculated by using Eq. \ref{equ:sigma}, for the 
states $0^+$, $2^+$, $4^+$, $6^+$, $8^+$, $10^+$ of $^{104,106,108,110}$Sn to see 
the goodness of the energy spectra before and after fitting processes.

Figure \ref{fig:Sn104_106} shows a comparison among the experimental, theoretical and
fitted spectra of the $^{104}$Sn isotope. Both fits give very similar energies for
the states $0^+$, $2^+$, $4^+$, $6^+$, $8^+$, $10^+$ of $^{104}$Sn. Both clonal 
selection and simplex fits improve the energies of the low-spin states $2^+$ and 
$4^+$ from 1.42 MeV to 1.31 MeV and 1.28 MeV, from 1.99 MeV to 1.90 MeV and 1.88 MeV, 
respectively, compare to the experimental data. The energy of the states 
$6^+$, $8^+$, $10^+$ is not improved through the both fits.

Figure \ref{fig:Sn104_106} also compares theoretical and fitted spectra of the 
$^{106}$Sn isotope to the experimental spectra. All fitted states merge to the 
experimental ones. The states $4^+$ and $10^+$ are exactly reproduced by simplex and 
the states $2^+$, $6^+$, $8^+$ are deviated slightly as to be 0.04, 0.04, 0.05 MeV 
and 0.02, 0.03, 0.05 MeV by clonal selection and simplex methods, respectively. 
Overall, the simplex methods gives slightly better spectrum with respect to the 
clonal selection method for the $^{106}$Sn isotope.

Figure \ref{fig:Sn108_110} shows the theoretical (before fitting) and experimental 
spectra of the $^{108}$Sn isotope as well as the fitted ones. From Table \ref{tab:rms},
the energy deviations of the states to the experimental ones are 0.11, 0.03 MeV for 
$2^+$, 0.12, 0.07 MeV for $4^+$, 0.01, 0.01 MeV for $6^+$, 0.09, 0.08 MeV for $8^+$, 
and 0.04, 0.00 MeV for $10^+$ with clonal selection and simplex methods, respectively.
As we see from these rms deviations, all states merge gradually to the corresponding 
experimental ones. This mergence shows a definite improvement on the spectrum of 
the $^{108}$Sn isotope.

Figure \ref{fig:Sn108_110} also shows the theoretical (before fitting) and experimental 
spectra of the  the $^{110}$Sn isotope as well as the fitted ones. All energy levels 
except the state $10^+$  merge to the experimental energies with the rms deviations of 
0.16, 0.06 MeV for $2^+$, 0.04, 0.06 MeV for $4^+$, 0.13, 0.14 MeV for $6^+$, 0.06,
0.05 MeV for $8^+$ of clonal selection and simplex methods, respectively. Both fit
processes do not improve the energy level of the state $10^+$. 

If we look at the systematic behavior of the states compare to before and after 
fitting processes, the states $2^+$ and $4^+$ are definitely improved with both fitting 
procedures for all $^{104,106,108,110}$Sn isotopes; the states $6^+$ and $8^+$ are 
fitted well for the $^{106,108,110}$Sn isotopes and not for $^{104}$Sn ; the state 
$10^+$ is fitted well for $^{106,108}$Sn isotopes and not for $^{104,110}$Sn 
isotopes.

To see the goodness of the obtained single-particle energies we have also carried 
out the shell-model calculations with the fitted single-particle energies for the 
lowest states $\frac{3}{2}^+$, $\frac{5}{2}^+$, $\frac{7}{2}^+$, $\frac{9}{2}^+$, 
$\frac{11}{2}^+$ of the light odd-even $^{103,105,107,109}$Sn isotopes.
Figures \ref{fig:Sn103_105} and \ref{fig:Sn107_109} show such a spectrum comparing
the calculated (before fitting) and available experimental values of the states
$\frac{3}{2}^+$, $\frac{5}{2}^+$, $\frac{7}{2}^+$, $\frac{9}{2}^+$, $\frac{11}{2}^+$ 
of the $^{103,105}$Sn and $^{107,109}$Sn isotopes, respectively, as well as the 
fitted ones. 

In Figure \ref{fig:Sn103_105}, the states $\frac{7}{2}^+$ and $\frac{11}{2}^+$ of 
$^{103}$Sn and $\frac{7}{2}^+$, $\frac{9}{2}^+$, $\frac{11}{2}^+$ of $^{105}$Sn are 
lowered well to get close to the experimental values. The states with higher energies
of $^{105}$Sn, e.g., $\frac{9}{2}^+$ and $\frac{11}{2}^+$, are fitted better with 
respect to the lower ones. 

In Figure \ref{fig:Sn107_109} the states $\frac{7}{2}^+$ of $^{107}$Sn and $^{109}$Sn 
are fitted to the values of 0.29 (0.27) and 0.08 (0.08) MeV with the rms deviations 
of 0.14 (0.12) MeV and 0.07 (0.07) MeV of clonal selection and simplex methods, 
respectively. The energy of $\frac{3}{2}^+$ state is much better fitted to the 
corresponding experimental value by lowering about -0.20 MeV. As in the case of 
$^{105}$Sn, the states with higher energies of $^{107}$Sn and $^{109}$Sn are fitted 
better with respect to the lower ones. 

As we see from the Figure \ref{fig:Sn103_105} and  \ref{fig:Sn107_109}, the fitted 
spectra of odd-even $^{103,105,107,109}$Sn isotopes agree with the experimental 
data better than the ones of even-even $^{104,106,108,110}$Sn isotopes. This is 
an evidence that the fitted single-particle energies give better spectrum for the 
other Sn isotopes which are not part of the fitting processes.

\section{Conclusion}

The non-linear fitting procedures, namely clonal selection and simplex methods, 
have been applied to optimize the single-particle energies in the $sdgh$ major 
shell in the nuclear shell model calculations. The fitted single-particle energies 
are calculated as to be $\varepsilon _{2s_{1/2}}=$1.53 (1.56) MeV, 
$\varepsilon _{1d_{3/2}}=$2.47 (3.20) MeV, and $\varepsilon _{0h_{11/2}}=$4.80 (4.15)
MeV with the simplex algorithm (the clonal selection algorithm). The fitted energy of
the single-particle state $2s_{1/2}$ are almost the same in both fit processes, but 
the other two fitted energies of the single-particle states are very different from 
each other. The optimization ideas of the simplex and clonal selection methods 
give this difference. The obtained fitted single-particle energies are still in 
the range of expectation. Indeed, there is a minimum and maximum limits of energy 
variables within the clonal selection algorithm while there is no limits of variables
within the simplex algorithm.

For the point of shell model calculations, we have repeated the shell model 
calculations of the states $0^+$, $2^+$, $4^+$, $6^+$, $8^+$, $10^+$ of the 
$^{104, 106,108,110}$Sn isotopes with the single-particle energies obtained 
before and after fitting processes to get better agreement of shell model spectra 
with the experiment. We have obtained better energy spectra of the $^{104,106,108,110}$Sn 
isotopes with the fitted single-particle energies. The fit processes improve the all states
of the $^{104,106,108,110}$Sn isotopes except the higher excited states of $^{104}$Sn
and the state $10^+$ of $^{110}$Sn. So, the overall agreement of the fitted spectra
is very good with experiment.

We have also performed the shell model calculations of the states $\frac{3}{2}^+$, 
$\frac{5}{2}^+$, $\frac{7}{2}^+$, $\frac{9}{2}^+$, $\frac{11}{2}^+$ of the light 
odd-even $^{103,105,107,109}$Sn isotopes with the single-particle energies obtained
before and after fitting processes. We have obtained even better energy spectra
for the $^{103,105,107,109}$Sn isotopes with respect to the $^{104,106,108,110}$Sn 
isotopes which are parts of the fitting processes.

The fitted single-particle energies with the simplex method give the spectra 
slightly better corresponding to the experiment with respect to the ones with clonal
selection method. The fitted single-particle energies can be used to do the shell
model calculations for the other nuclei in the vicinity of $^{100}$Sn, especially
for mid-heavy Tin (Sn) isotopes with $A=110-120$ and light Antimony (Sb) isotopes.
With this study, we present an improvement on the ambiguity of the single-particle
energies to be used in future shell model calculations in the $sdgh$ major shell.

%%%%%%%%%%%%%%%% acknowledgments %%%%%%%%%%%%%%%%%
\begin{acknowledgments}
This work was supported in part by S\"{u}leyman Demirel University under 
Contract No. SDUBAP 1822-YL-09 and 
The Scientific and Technological Council of Turkey under Contract No. TUBITAK 105T092.
\end{acknowledgments}

%%%%%%%%%%%%%%%% bibliography %%%%%%%%%%%%%%%%%%%%

%New pages for tables and figures

\end{document}